\documentclass[10pt,conference,final]{IEEEtran}
\IEEEoverridecommandlockouts
\usepackage{cite}
\usepackage{amsmath,amssymb,amsfonts}
\usepackage{algorithmic}
\usepackage{graphicx}
\usepackage{subfigure}
\usepackage{textcomp}
\usepackage[table]{xcolor}
\usepackage{booktabs}
\usepackage{epstopdf}
\usepackage{multirow}
\usepackage{array}
\usepackage{makecell}
\usepackage{enumerate}
\def\BibTeX{{\rm B\kern-.05em{\sc i\kern-.025em b}\kern-.08em
    T\kern-.1667em\lower.7ex\hbox{E}\kern-.125emX}}
\begin{document}

\title{A Two-Dimensional Deep Network for RF-based Drone Detection and Identification Towards Secure Coverage Extension\\
	\thanks{This work reported in this paper is supported in part by the National Key R\&D Program of China under Grant No. 2020YFB1806905, the National Natural Science Foundation of China under the Grant No. 62071368, and the Key Research and Development Program of Shaanxi Province under the Grant No. 2023-GHZD-05.}}

\author{
	\IEEEauthorblockN{\normalsize Zixiao Zhao$^{1,2} $, Qinghe Du$^{1,2} $, Xiang Yao$^{1,2} $, Lei Lu$^{1,2} $, and Shijiao Zhang$^{1,2} $
		\\
		1 School of Information and Communications Engineering, Xi'an Jiaotong University, China
		\\
		2 National Simulation Education Center for Communications and Information Systems, Xi'an Jiaotong University, China
		\\
		Emails: \{\textit{zzx67120787@stu.xjtu.edu.cn, duqinghe@mail.xjtu.edu.cn},
		\\
		\textit{ 630321@stu.xjtu.edu.cn, lu.lei@xjtu.edu.cn, zhshijiao@mail.xjtu.edu.cn}\} 
}}

\maketitle

\begin{abstract}
As drones become increasingly prevalent in human life, they also raises security concerns such as unauthorized access and control, as well as collisions and interference with manned aircraft.
Therefore, ensuring the ability to accurately detect and identify between different drones holds significant implications for coverage extension. 
Assisted by machine learning, radio frequency (RF) detection can recognize the type and flight mode of drones based on the sampled drone signals.
In this paper, we first utilize Short-Time Fourier Transform (STFT) to extract two-dimensional features from the raw signals, which contain both time-domain and frequency-domain information. Then, we employ a Convolutional Neural Network (CNN) built with ResNet structure to achieve multi-class classifications.
Our experimental results show that the proposed ResNet-STFT can achieve higher accuracy and faster convergence on the extended dataset. Additionally, it exhibits balanced performance compared to other baselines on the raw dataset.

\end{abstract}


\section{Introduction}
Nowadays, the applications of drones, which are also known as unmanned aerial vehicles (UAVs), have been penetrated into every aspects of human life, including aerial photography, plant protection, military, etc. 
The global civil drone industry is expected to reach about 21.6 billion U.S. dollars by 2027. Military use has previously accounted for much of drone use, but the industry is increasingly entering commercial, scientific, and agricultural usage.

While drones offer numerous benefits and opportunities, they also present several security concerns that need to be addressed \cite{b1}, raising challenges for coverage extension.
One of the primary concerns is the potential for unauthorized access and control of drones. Hackers or malicious individuals could attempt to gain control over a drone by exploiting vulnerabilities in its communication systems or flight controls. This can lead to misuse of the drone for illegal activities or sabotage.
Another concern arises from the growing number of drones in the airspace, which increases the risk of collisions and interference with manned aircraft. Unauthorized or unregulated drone flights can pose risks to aviation safety, especially near airports or in restricted airspace.
An example occurred in Oct. 2017 that a civil aircraft collided with a drone as the former was approaching the airport near Quebec City, Canada.
Lastly, data breaches and privacy issues are also hidden troubles. Drones often capture and transmit data during their operations, including images, videos, and other sensor readings. If proper security measures are not in place, there is a risk of data breaches, where the captured information can be intercepted or accessed by unauthorized parties. This raises concerns about privacy violations and the potential misuse of sensitive data \cite{b2}.

Thus as drones become increasingly prevalent, ensuring the ability to accurately detect and identify between different drones holds significant implications for secure coverage extension. 
 In order to reduce or eliminate the threats posed by illegal drone flights, there are four main detection methods: optical detection, acoustic detection, radar detection, and radio frequency (RF) detection \cite{b3}.
 Compared with the other three detection methods, the advantages of RF detection which will be achieved based on the captured communication signals include:
 it can detect drones of any size and distance, i.e., within line of sight or beyond line of sight, and can also be used to identify the flight mode of drones, such as flying, hovering, recording, etc \cite{b4}.
 
 In this paper, we design a deep network to accurately detect the presence of drone signals and identify various drone states, taking into full consideration the characteristics of sampled radio signals.
 The contributions of this work include:
\begin{enumerate}[1)]
	\item We extend the raw dataset considering the situation that several different types of drones are coexisting and simultaneously transmitting signals. Correspondingly, we design extra classification tasks which have not been discussed in previous research. 
	\item We employ short-time fourier transform (STFT) algorithm to extract two-dimensional features, i.e., time-domain and frequency-domain. They can provide more information of hidden correlations for classification by feeding into two-dimensional convolutional neural network (2D CNN).
	\item Our experiments show that the proposed ResNet-STFT algorithm is able to achieve 98.7\% accuracy in seven-class classification based on the extended dataset. Moreover, it can also achieve faster convergence compared with the one-dimensional baseline method.
\end{enumerate}
 
 The rest of the paper is organized as follows. Section II introduces the related works. Section III describes the dataset and extended version. Section IV proposes our methodology including feature extraction and classification network. Section V presents the experimental results. Finally, Section VI concludes the work.
 
\section{Related Works}
Machine learning has been widely used in drone detection.
Nie. et al. in \cite{b3} extracted fractal dimension, axially integrated bispectra, and square integrated bispectra as UAV radio frequency (RF) fingerprint.
The principal component analysis (PCA) algorithm was applied to reduce the dimensionality of features, then machine learning classifier achieved UAV identification.
Medaiyese. et al. in \cite{b4} proposed a three-level hierarchical framework to detect UAV signals in the presence of other wireless signals such as Bluetooth and WiFi, which utilized a semi-supervised learning approach.

DroneRF is a common drone dataset proposed by Allahham. et al. in \cite{b5}.
Some research have been done based on this dataset.
In \cite{b6} authors proposed a fully connected neural network with three hidden layers to classify drone signals.
Allahham. et al. in \cite{b7} further proposed a multi-channel 1D CNN achieving higher accuracy performances.
In \cite{b8} Raina. et al. proposed ConvLGBM model which combined the feature extraction capability of a CNN network with the high classification accuracy of the Light Gradient Boosting Machine (LightGBM). 
Considering feature engineering, Inani. et al. in \cite{b9} synthetically discussed several features in time and frequency-domain, including root mean square energy (RMS), discrete-fourier transform (DFT), power spectral density (DFT), etc. 
They further proposed a 1D CNN to identify the target drone signals with these extracted features.
In this paper, we will extend the raw DroneRF dataset, and introduce feature extraction methods and classification networks working in two-dimensional space.

\section{Dataset and Extended Version}
\subsection{DroneRF dataset}

M.S.Allahham et al. in \cite{b5} provide the DroneRF dataset which they collected from three types of drones and five types of function modes.
Specifically, Parrot Bebop and Parrot AR Drone were both tested in Off, On and connected, Hovering, Flying, and Video recording modes. Another DJI phantom was merely tested in two modes, i.e., Off and On and connected. 
The RF receiver can capture the transmission signals between the drone and controller, which sampling bandwidth is equal to 40MHz.
Thus for scanning 80MHz spectrum, the authors adopted two RF receivers, sampling for the lower 0$ \sim $40MHz frequency band and higher 40$ \sim $80MHz frequency band separately.
Each sample of this dataset, which is also denoted as a \textit{segment}, is composed of $ 10^{7} $ time-domain points.
The dataset has 227 segments in total, and the proportions of them can be listed in Table \ref{dronerf}.

Moreover, the authors design binary unique identifier (BUI) rule to effectively name and distinguish the segments.
The BUI number of each segment is composed of five binary digits.
The first number indicates the presence of drone activities, the second and third numbers are used to characterize the three drone types, and the last two digits are corresponding to the four function modes.
This scientific naming rule will also be practiced in the following dataset augmentation.
In Table \ref{dronerf} the samples with BUI belonging to 0xxxx or 1xxxx come from the raw dataset.
Others in Table \ref{dronerf} are extended samples, and we will introduce them below.

\subsection{Data augmentation}
We further consider the extended condition that there are two types of drones coexisting and working in the same mode in a segment.
Due to the records of Phantom merely contain On and connected mode in the raw DroneRF dataset, we mainly discuss this mode.
Specifically, we add the time-domain points of Bebop \& AR, Bebop \& Phantom, and AR \& Phantom respectively, indicating the two drones transmitting signals simultaneously.
We show the detailed information in Table \ref{dronerf}.

Moreover, we utilize BUI 2xxxx to name the extended data.
The first number indicates there two types of drones are coexisting.
The second and third numbers, i.e., 00, 01, and 10, are related to Bebop \& AR, Bebop \& Phantom, and AR \& Phantom, respectively.
The last two numbers indicates the function mode, which are merely set as 00 because of On and connected mode.
\begin{table}[]
	\centering
	\renewcommand\arraystretch{1.15}
	\caption{Summary of DroneRF Dataset and its Extended Version}
	\begin{tabular}{c|cccc}
		\hline \hline
	BUI	&    Types   & Function modes &  \makebox[0.05\textwidth][c]{\makecell{Numbers \\of samples}} & No.\\
	\hline
	0xxxx & No drone & No drone & 41 & 00000\\
	\hline
		\multirow{9}{*}{1xxxx} & \multirow{4}{*}{Bebop}  & On and connected &  21 & 10000\\
													&                   						& Hovering &  21 & 10001\\
													&                  				 			& Flying &  21 & 10010\\
													&                  				 			& Video recording&  21 & 10011\\
													\cline{2-5}
													& \multirow{4}{*}{AR}	 & On and connected &  21 & 10100\\
													&                  						 & Hovering &  21 & 10101\\
													&                  						 & Flying &  21 & 10110\\
													&                   					& Video recording &  18 & 10111\\
													\cline{2-5}
													&           Phantom             & On and connected & 21 & 11000\\
	\hline
	 \multirow{3}{*}{2xxxx} & Bebop \& AR & On and connected & 441  & 20000\\
	 											& Bebop \& Phantom & On and connected & 441  & 20100\\
	 											& AR \& Phantom & On and connected & 441  & 21000\\
	\hline \hline
	\end{tabular}
\label{dronerf}
\end{table}

\subsection{Classification cases}
\label{Section-Classification cases}
According to the data types of dataset, we design five kinds of classification cases described as follows:
\begin{itemize}
	\item Case I: Binary classification. The classifier needs to detect whether a piece of given data contains drone signals.
	\item Case II-A: Four-class classification. The classifier needs to identify none or which type of drone signal a piece of given data contains, including Bebop, AR, or Phantom.
	\item Case II-B: Three-class classification. The classifier needs to identify which two types of drone signals a piece of given data contains, including Bebop \& AR, Bebop \& Phantom, or AR \& Phantom.
	\item Case II-C: Seven-class classification. The integrated case of Case II-A and Case II-B.
	\item Case III: Ten-class classification. The classifier needs to identify none or which type of drone signal and its function mode a piece of given data contains.
\end{itemize}

Note that Case I, Case II-A, and Case III are based on the raw DroneRF dataset which have been discussed by other papers \cite{b6,b7,b8,b9}, while Case II-B and Case II-C are based on our extended dataset.
The following experiments will compare our detection algorithm with other outstanding baselines and prove that ours can achieve balanced performances on general cases and better performances on extended cases.

\section{methodology}
\subsection{Feature extraction with Short-time fourier transform}
\label{Section-STFT}
In \cite{b6} a segment was divided into several continuous but non-overlapping time-domain parts, and we denote this as simple-cutting (SCU) method.
Compared with SCU, short-time fourier transform (STFT) algorithm compensates the information loss by introducing window function.
Specifically, the overlapping part between two windows contributes to catch hidden but continuous time-domain information.
The formulation of STFT can be depicted as follows:

\begin{equation}
	STFT(\tau,f) = \int_{-\infty}^{+\infty}x(t)h(t-\tau)e^{-j2\pi ft}d\tau,
\end{equation}
where $ x(t) $ denotes the target signal, and $ h(t-\tau) $ denotes a window function which is used for intercepting a \textit{frame} from $ x(t) $.
Shifting $\tau $ along timeline, STFT can get the fourier transform result of each intercepted frame, and finally, help to analyze the whole time-domain and frequency-domain information of $ x(t) $.

The function \textit{spectrogram(x,window,noverlap,nfft,fs)} in MATLAB can fulfill the calculation of STFT.
We set the parameter \textit{nfft} as 128, indicating 128-point FFT for each frame.
The type and length of window function both can be altered.
To ensure that the numbers of time-domain and frequency-domain points are both equal to 128, which will be convenient for the following CNN to recognize and classify, we set the length of \textit{Hamming} window as $ 8.8\times10^4 $ and overlapping parts between two windows as $ 10^4 $ accordingly.
The overlapping ratio is equal to 11.4\%.

\begin{figure}
	\centering
	\subfigure[]{		
		\includegraphics[width=4.2cm]{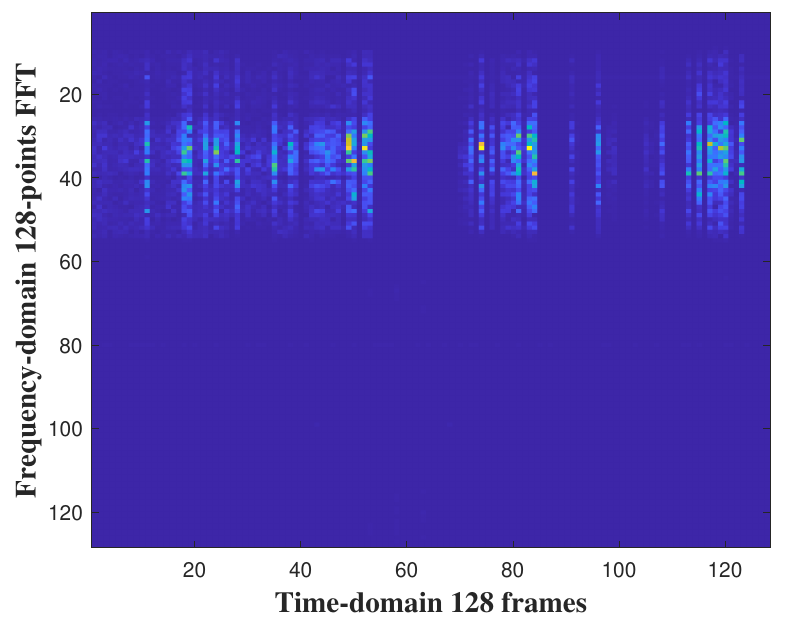}
		\label{STFT-20000}
	}%
	\subfigure[]{
		\includegraphics[width=4.2cm]{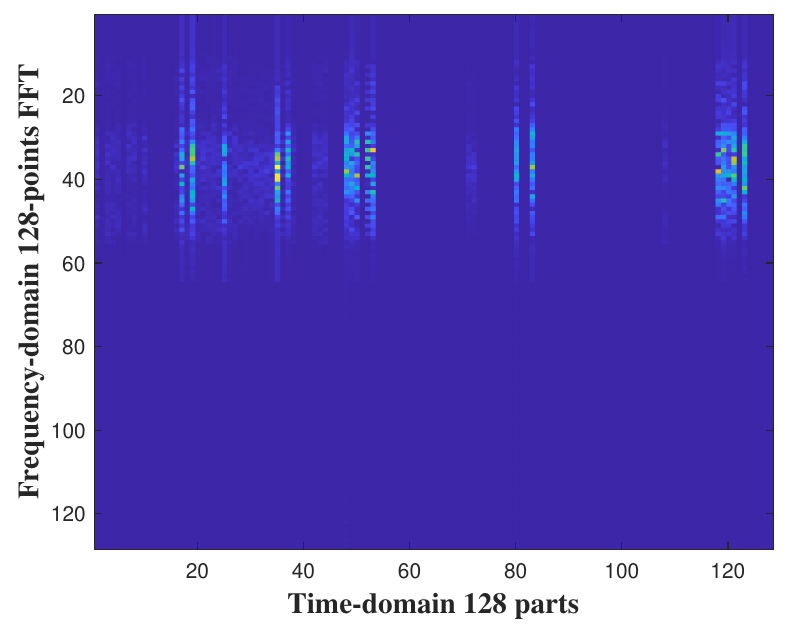} 
		\label{SCU-20000}
	}
	\centering
	\caption{These figures are spectrogram of signals when Bebop drone and AR drone are coexisting. The size is equal to 128*128. The horizontal axis indicates time-domain containing 128 intercepted parts, and the vertical axis indicates frequency-domain containing 128-points FFT. (a) is generated with STFT algorithm, and (b) is generated with SCU method. Compared with (b), some hidden features are exhibited in (a).}
	\label{STFT and SCU}
\end{figure}

We further compare the STFT algorithm with SCU method in visualization form.
The upper-half spectrogram is generated with lower sampling band, while the lower-half spectrogram is generated with higher sampling band that were sampled simultaneously.
Hereinafter, we will use this concatenation way to generate feature patterns from dataset.
For instance, we choose a segment coming from BUI = 20000 which indicates that Bebop drone and AR drone are coexisting.
As shown in Fig. \ref{STFT and SCU}, compared with SCU style (Fig. \ref{SCU-20000}), the frequency points with high energy in STFT style (Fig. \ref{STFT-20000}) tend to be more concentrative, and some hidden features are also enhanced. We believe that STFT is able to provide additional and valuable information for identification.

Moreover, it is evident that high energy points mainly appear in the upper-half spectrogram. 
Authors in \cite{b10} discussed the performances with lower band, higher band, and both two bands, respectively. They believed that the lower sampling band has carried enough features for detecting and identification drones.
Our spectrogram results confirm this conclusion well.

\subsection{ResNet structure}
Convolutional neural network (CNN) is one of the most typical algorithm in deep learning, which contains convolutional computation and deep structure.
The applications of CNN are mainly related to computer vision area, such as image classification, semantic segmentation, pose estimation, etc.
Especially when the dimensions of input data are quite large, CNN is able to avoid explosive scale of network parameters by local and distributed convolutional computing.

One-dimensional (1D) CNN is typically used for signal processing, in which the input data contains correlation characteristic on timeline and needs to be predicted or classified.
Two-dimensional (2D) CNN has border application prospect, in which the input data is shaped into matrix form.
The typical 2D CNN structures include AlexNet \cite{b11}, VGG \cite{b12}, GoogleNet \cite{b13}, ResNet \cite{b14}, etc.
In this paper we employ 2D CNN with ResNet structure to solve the problems.

In computer vision, the depth of the network will increase with the larger scale of input features.
Typically deeper network is able to achieve better performances, however, the new problem, i.e., gradient explosion and gradient vanishing, which will result in the failure of network convergence, becomes an obstacle to training such a network.
ResNet structure is proposed to counter this problem.
Firstly, the shortcut connection of ResNet structure can accelerate the information propagation in the whole network.
Secondly, the batch normalization (BN) is utilized to ensure that the input feature map of every convolutional layer follows the normal distribution with mean 0 and variance 1.
The formulations of BN can be depicted as follows \cite{b15}:
\begin{equation}
	\left\{
	\begin{split}		
		&y_i = \gamma \hat{x_i} + \beta ;\\
		&\hat{x_i} = \frac{x_i -\mu_{\mathcal{B}} }{\sqrt{\sigma_{\mathcal{B}}^2 + \epsilon}}; \\
		&\sigma_{\mathcal{B}}^2  = \frac{1}{m} \sum_{i=1}^{m} (x_i - \mu_{\mathcal{B}} )^2; \\
		&\mu_{\mathcal{B}} = \frac{1}{m} \sum_{i=1}^{m}  x_i.
	\end{split}
	\right.\label{Equ-Resnet}
\end{equation}

The input of BN is a mini-batch $ \mathcal{B} = \{x_{1 \dots m}\} $, and for each value $ x_i $ of $  \mathcal{B}$, the output is denoted as $ y_i = \text{BN}_{\gamma,\beta}(x_i) $.
The $ \mu_{\mathcal{B}}  $ and $ \sigma_{\mathcal{B}}^2  $ represent the mean and variance value of mini-batch, respectively.
The $ \hat{x_i} $ is the result of normalization.
Moreover, the parameter $ \gamma $ and $ \beta $ need to be learned in the back propagation process, which reflect the mean and variance of the whole training dataset and can help to scale and shift the normalized $ \hat{x_i} $.

\subsection{ResNet-CNN classifier for STFT feature map identification}

\begin{figure}
	\centering
	\subfigure[]{		
		\includegraphics[width=8.5cm]{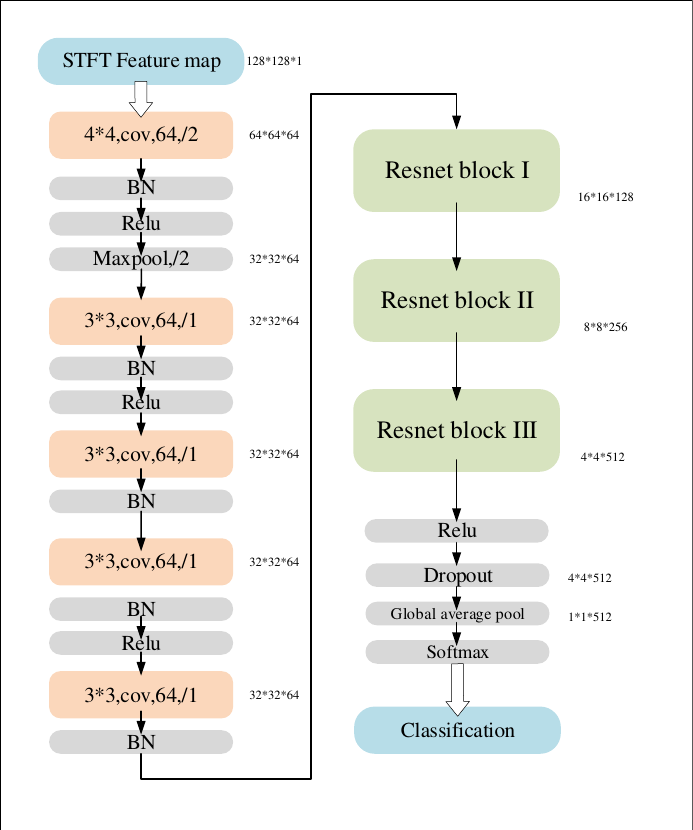}
		\label{72Layers-Resne}
	}%
		
	\subfigure[]{
		\includegraphics[width=7.5cm]{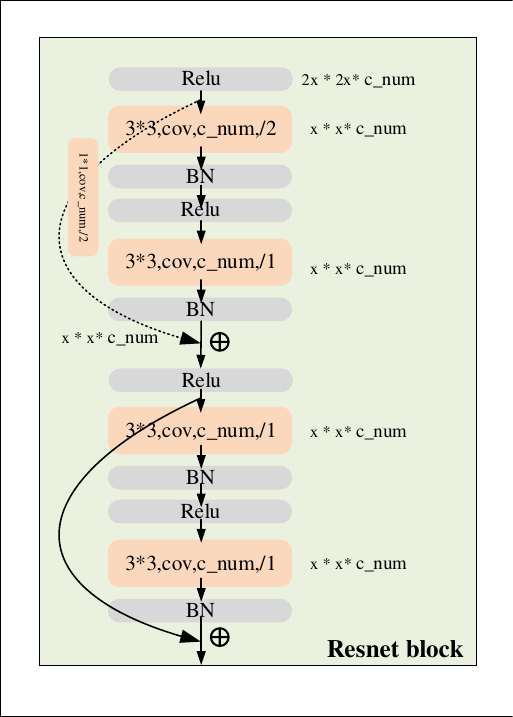} 
		\label{Resnet-block}
	}
	\centering
	\caption{The network structure of ResNet-CNN. (a) shows the complete structure, where the input is STFT feature map and the output is classification result. (b) shows the internal structure of Resnet block which decreases the height and width dimensions of input but increases the depth dimension.
}
	\label{Resnet}
\end{figure}

In this paper we develop a 72-layers ResNet, as shown in Fig. \ref{Resnet}.
The input of network is 128*128 STFT feature map which has been introduced in Section \ref{Section-STFT}.
For each orange rectangle, which represents the convolutional layer, the first number denotes the size of filter, the second denotes the number of filters, and the last number denotes stride.
The internal structure of Resnet block is shown in Fig. \ref{Resnet-block}.
For Resnet block I, II, and III, the numbers of filters, i.e., the values of parameter $ c\_num $, are equal to 128, 256, and 512, respectively.
Note that the shortcut connection of the upper-half part includes dimension reduction implemented by a convolutional layer (stride=2).
Lastly, the traditional fully connected layer is replaced with the global average pooling layer, thus the feature map is directly fed into the softmax layer, assisting to reduce network parameters.

\section{Experiment Results}
\subsection{Experimental setup}

\subsubsection{Baselines}
We employ the newly proposed work in \cite{b9} as baseline which has been verified to achieve good accuracy performances on the raw DroneRF dataset.
It developed a 1D-CNN network and recommended to extract power spectral density (PSD) features for the raw signals.
For brief descriptions, we simply denote this baseline as 1D-PSD, and denote our recommended algorithm as ResNet-STFT below.
The formulation of PSD is as follows, where $ x_N $ denotes $ N $ sampling points:
\begin{equation}
	P(w) = \frac{1}{N}\left\|\sum_{n=0}^{N-1}x_N(n)e^{-jwn}\right\|^2.
\end{equation}


\subsubsection{Metrics}
We use the common metrics, i.e., accuracy, precision, recall, and F1-score, to evaluate the classification performances. 
Especially, for the multi-class classification, we calculate `one vs rest' separately for each class and choose the average value to denote the overall performance.
Moreover, we will also show the confusion matrix of each classifier.
The formulations can be depicted as follows:
\begin{equation}
	\begin{split}
		\text{Accuracy} &= \frac{TP+TN}{TP+TN+FP+FN};\\
		\text{Precision} &= \frac{TP}{TP+FP};\\
		\text{Recall} &= \frac{TP}{P};\\
		\text{F1-score} &= \frac{2*\text{Precision}*\text{Recall}}{\text{Precision}+\text{Recall}}
	\end{split}
\end{equation}

\subsubsection{Experiment settings}
The experiments have been carried out on MATLAB.
STFT feature maps are extracted from the raw data by \textit{spectrogram} function.
We use \textit{deepNetworkDesigner} toolbox to design and analyze the deep networks.
The network training function is \textit{trainNetwork}, where we set: optimizer$ -> $`\textit{adam}', initial learn rate$ -> $0.0001, and L2 regularization factor$ -> $0.0001.
Considering different scales of cases which have been defined in Section \ref{Section-Classification cases}, for the experiments based on the raw dataset, we set `MiniBatchSize'=8 and `MaxEpochs'=50, while for those based on the extended dataset, we set `MiniBatchSize'=32 and `MaxEpochs'=5.
The loss function is \textit{cross entropy.}

We repetitively carry out each experiment for five times, and use the average as the final results.
In each training-testing process, we divide the whole dataset into three sections, i.e., 80\% training data, 10\% validation data, and 10\% testing data.

\subsection{Experiment I: Classifications on raw dataset}
This experiment compares the performances of our proposed ResNet-STFT with baseline 1D-PSD on Case I, Case II-A, and Case III.
Specifically, the cases are related to two, four, and ten-class classifications, respectively.
The results can be found in Table \ref{Table-exp1}.

In binary classification, two models both can achieve no error, i.e., always accurately detect whether the sample contains drone signals.
In complex multi-class classifications, there comes some decline on accuracy, especially in ten-class classification.
The factors that contribute to this situation include: 1) the small number of samples in each class is not sufficient for training; 2) correspondingly, the overfitting problem appears in deep network which has significant affects on the testing dataset.
In the future work, we will further develop DroneRF dataset, such as adding noise, to compensate this problem.
\begin{table}[htbp]
		\renewcommand\arraystretch{1.15}
		\centering
	\caption{Accuracy and F1-score of Two, Four, and Ten-Class classifications}
	\begin{tabular}{c|cc|cc|c}
		\hline\hline
		& \multicolumn{2}{c}{Case I} & \multicolumn{2}{c}{Case II-A} & Case III\\
		\cline{2-6}
		&     Accu      &     \makebox[0.05\textwidth][c]{F1-score}     &       Accu    &     \makebox[0.05\textwidth][c]{F1-score}      &      Accu      \\
		\hline
ResNet-STFT		&     1      &     1     &      0.985     &     1     &        0.682     \\
\hline
1D-PSD		&     1      &     1     &      0.955     &     0.98     &     0.697       \\
\hline\hline
	\end{tabular} \label{Table-exp1}
\end{table}

\subsection{Experiment II: classifications on extended dataset}
This experiment compares the performances of our proposed ResNet-STFT with baseline 1D-PSD on Case II-B and Case II-C.
Specifically, the cases are related to three and seven-class classifications, respectively.
The results can be found in Table \ref{Table-exp2}.

\begin{figure}
	\centering
	\subfigure[]{		
		\includegraphics[width=7cm]{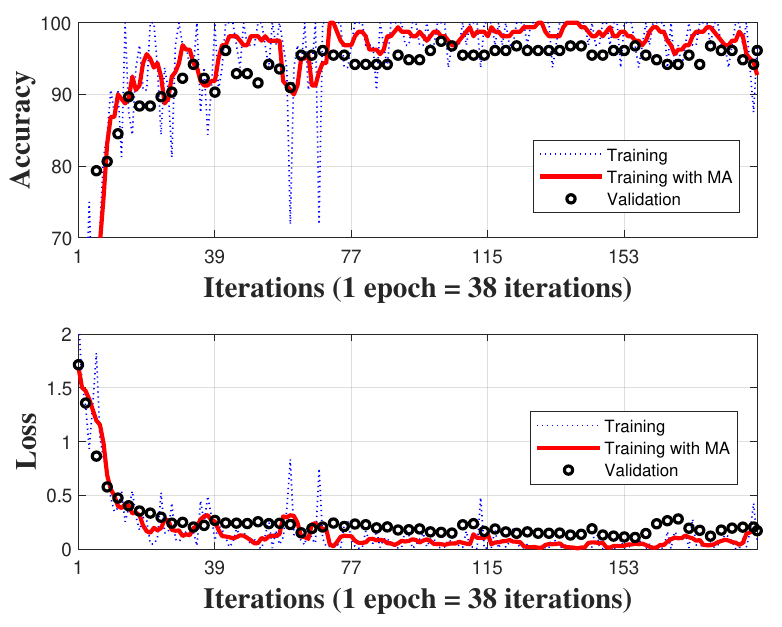}
		\label{Seven-Res-train}
	}%
	
	\subfigure[]{
		\includegraphics[width=7cm]{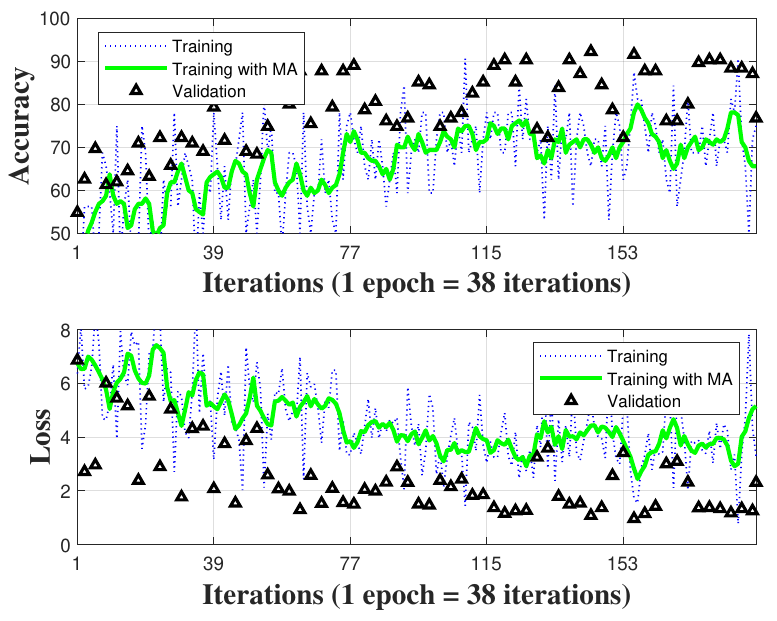} 
		\label{Seven-PSD-train}
	}
	\centering
	\caption{The loss and accuracy curves of training and validation process. (a) shows performances of ResNet -STFT, and (b) shows performances of 1D-PSD. `Training with MA' denotes the moving average (MA) results of every five Training points. 
		There are 190 iterations contributed to 5 epochs in total.}
	\label{Seven-train-val}
\end{figure}

In Case II-B, our network can precisely identify which two types of drones are coexisting.
The accuracy and f-score performances of ResNet-STFT are a bit better than those of 1D-PSD.
In Case II-C, the detection algorithm needs to identify which type(s) of drones it contains, i.e., none, single three types, or coexisting in pairs.
The accuracy of ResNet-STFT can achieve 98.7\%, which also takes less time to converge.
As shown in Fig. \ref{Seven-Res-train}, the accuracy and loss curves of ResNet-STFT quickly converge into relatively ideal levels in the first epoch.
Comparatively, after 5 epochs 1D-CNN still cannot well converge and merely reaches 81.1\% accuracy.

\begin{figure}
	\centering
	\subfigure[]{		
		\includegraphics[width=4cm]{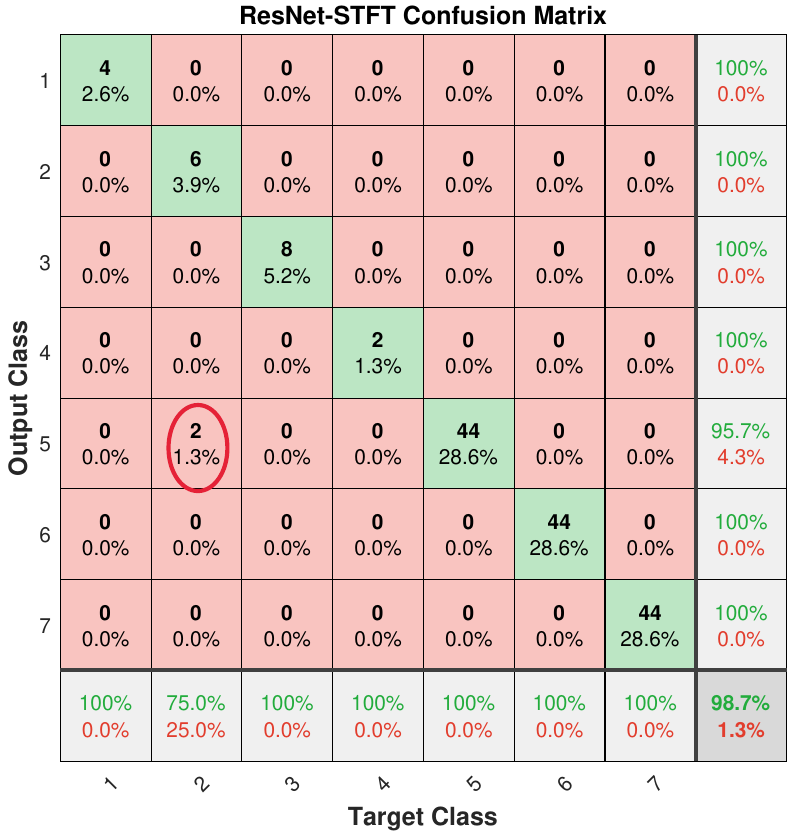}
		\label{Seven-Res-confusion}
	}%
	\subfigure[]{
		\includegraphics[width=4cm]{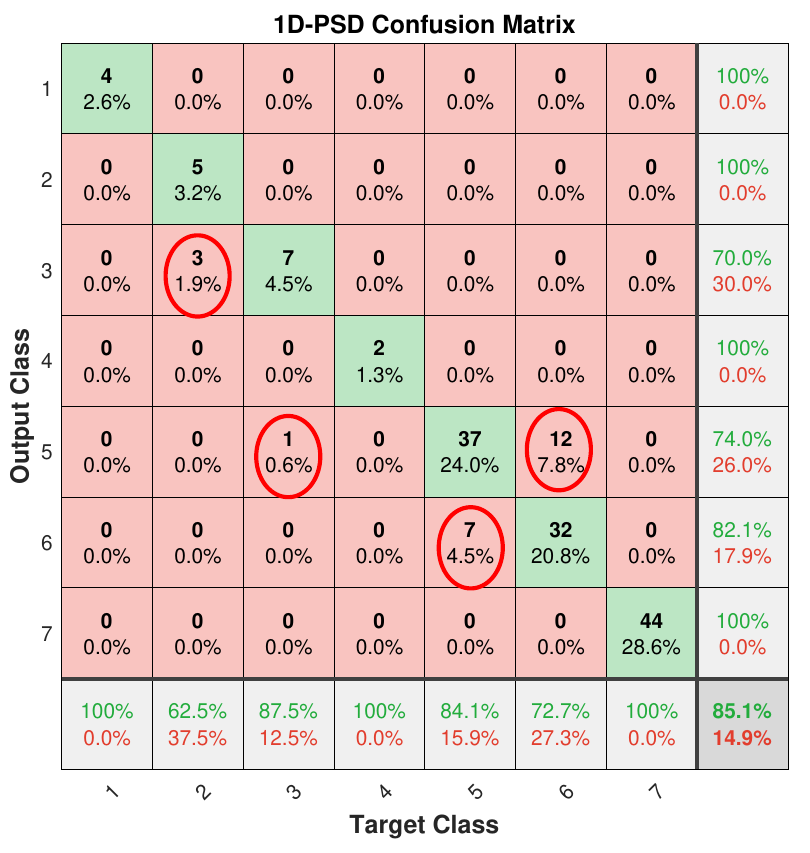} 
		\label{Seven-PSD-confusion}
	}
	\centering
	\caption{The confusion matrix of ResNet-STFT (a) and 1D-PSD (b). 
	The column on the far right of the figure shows the percentages of all the examples predicted to belong to each class that are correctly classified, i.e., precision. 
	The row at the bottom of the figure shows the percentages of all the examples belonging to each class that are correctly classified, i.e., recall.
	The cell in the bottom right of the figure shows the overall accuracy.}
	\label{Seven-confusion}
\end{figure}

Moreover, we also show the confusion matrix figures in Fig. \ref{Seven-confusion}.
ResNet-STFT mistook two samples of class 2  into class 5,  which represent single Bebop and Bebop \& AR respectively.
However, 1D-PSD seriously confused class 5 and class 6, which represent Bebop \& AR and Bebop \& Phantom respectively. The precision of class 5 and class 6 are equal to 74\% and 82.1\%.

Thus ResNet-STFT has potential of identifying complex energy signals.
On the one hand, STFT features catch more time-domain information which can perform as a good supplement to frequency-domain.
Especially when several signals overlap, excavating the correlations between time-domain and frequency-domain will be significant.
On the other hand, ResNet-STFT is much deeper than 1D-PSD, but its scale of parameters does not explosively increase.
On the contrary, it can achieve faster convergence compared with baseline.

\begin{table}[htbp]
	\centering
	\renewcommand\arraystretch{1.15}
	\caption{Accuracy and F1-score of Three and Seven-Class classifications}
	\begin{tabular}{c|cc|cc}
		\hline\hline
		& \multicolumn{2}{c}{Case II-B} & \multicolumn{2}{c}{Case II-C}\\
		\cline{2-5}
		&     Accuracy      &     F1-score     &       Accuracy    &    F1-score   \\
		\hline
		ResNet-STFT		&     1      &     1     &     0.987     &     1    \\
		\hline
		1D-PSD		&     0.947      &     0.98    &      0.811     &     0.924 \\
		\hline\hline
	\end{tabular} \label{Table-exp2}
\end{table}

\section{Conclusions}
As drones become increasingly prevalent in human life, ensuring the ability to accurately detect and identify between different drones holds significant implications for public safety. 
We selected a common dataset DroneRF to verify our algorithm.
Moreover, based on the raw dataset, we also considered the extended condition that there were two types of drones coexisting.
We first utilized Short-Time Fourier Transform (STFT) to extract two-dimensional features from the raw signals, which contained both time-domain and frequency-domain information. Then, we employed a Convolutional Neural Network (CNN) built with ResNet structure to achieve multi-class classifications.
Our experimental results showed that the proposed ResNet-STFT could achieve higher accuracy and faster convergence on the extended dataset, especially 98.7\% accuracy in seven-class classification.
Additionally, it exhibited balanced performance compared to other baselines on the raw dataset.

In the future, we will further develop DroneRF dataset, such as adding noise, to compensate the problem that too small numbers of samples can be trained in ten-class classification.
Moreover, we also consider employ and cascade other efficient machine learning classifiers after extracting abstract features by our ResNet-STFT.

\vspace{12pt}
\color{red}

\end{document}